\documentclass[a4paper,11pt]{article}
\pdfoutput=1 % if your are submitting a pdflatex (i.e. if you have
\usepackage{jinstpub} % for details on the use of the package, please
\title{A Simple Modification of DC Current Septa to Reduce Current Density by Half}

\author{Jay Benesch}
\affiliation{Thomas Jefferson National Accelerator Facility,\\Newport News, VA, USA}

\emailAdd{benesch@jlab.org}

\abstract{
Circular accelerators typically have one injection and one extraction septum magnet \cite{1}.  CEBAF is a recircuating electron linac which has a total of 27 DC current septa and one Lambertson. \cite{2}  Current densities range from 28-48 $A/mm^2$.  Current sheet widths are 5-24 mm, turns count 5-24 and lengths 1000-3000 mm.  A design exercise to increase the beam energy to 22 GeV is underway. \cite{10} Since doubling the current density in copper is not practical in the CEBAF layout a conductively cooled superconducting septum concept was examined.  Putting the current sheet and its cryostat between the poles as in a standard current sheet septum would have required 90 mm pole gap.  The poles were brought to 40 mm separation and the steel notched for the 90 mm cryostat.  The field in the bore increased while the field outside the current sheet remained close to zero as in a conventional septum.  Required current density dropped enough that a copper coil became possible.  Two examples will be shown, the one discussed above and the modification of the 3000 mm septum with 0.92 T bore field. 
}

\keywords{Instrumentation and hardware for accelerators}

\arxivnumber{} % only if you have one

\begin{document}
\maketitle
\flushbottom

\section{Previous work}
\label{sec:intro}
The variety of septa used in accelerators are discussed in \cite{1}.  These include direct drive DC septum magnet, direct drive pulsed septum magnet, eddy-current septum and Lambertson septum.  In this work we discuss a simple modification only of the first type, DC septum.  In Figure 1 an idealized model of the thinnest septum used in CEBAF is shown.  There is a thin current sheet with 30 $A/mm^2$ at the left and a conductor with five times the area at the right.  This style of DC septum requires great care in designing the cooling of the current sheet.  A recent example with 105 $A/mm^2$ in the copper and DC current 5812 A is discussed in \cite{3}. 
\\
\graphicspath{{Images/}}
\begin{figure}[htp]
    \centering
    \includegraphics[width=8cm]{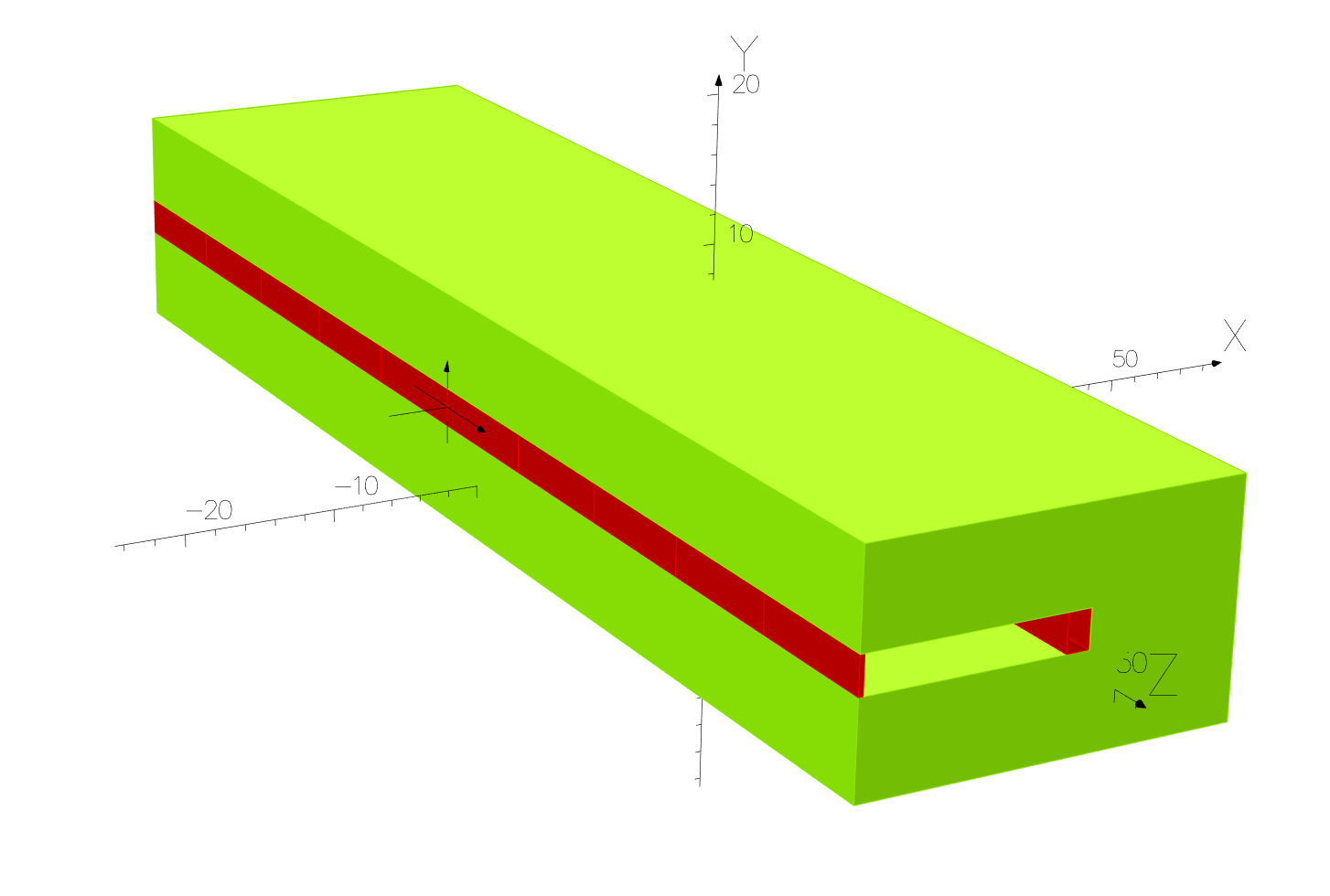}
    \caption{Thinnest current sheet septum magnet used at CEBAF.  The beams on either side of the current sheet are in the same vacuum space as the coil.  All other CEBAF septa have separate vacuum chambers for passing and deflected beams.  }
    \end{figure}

\graphicspath{{Images/}}
\begin{figure}[htp]
    \centering
    \includegraphics[width=8cm]{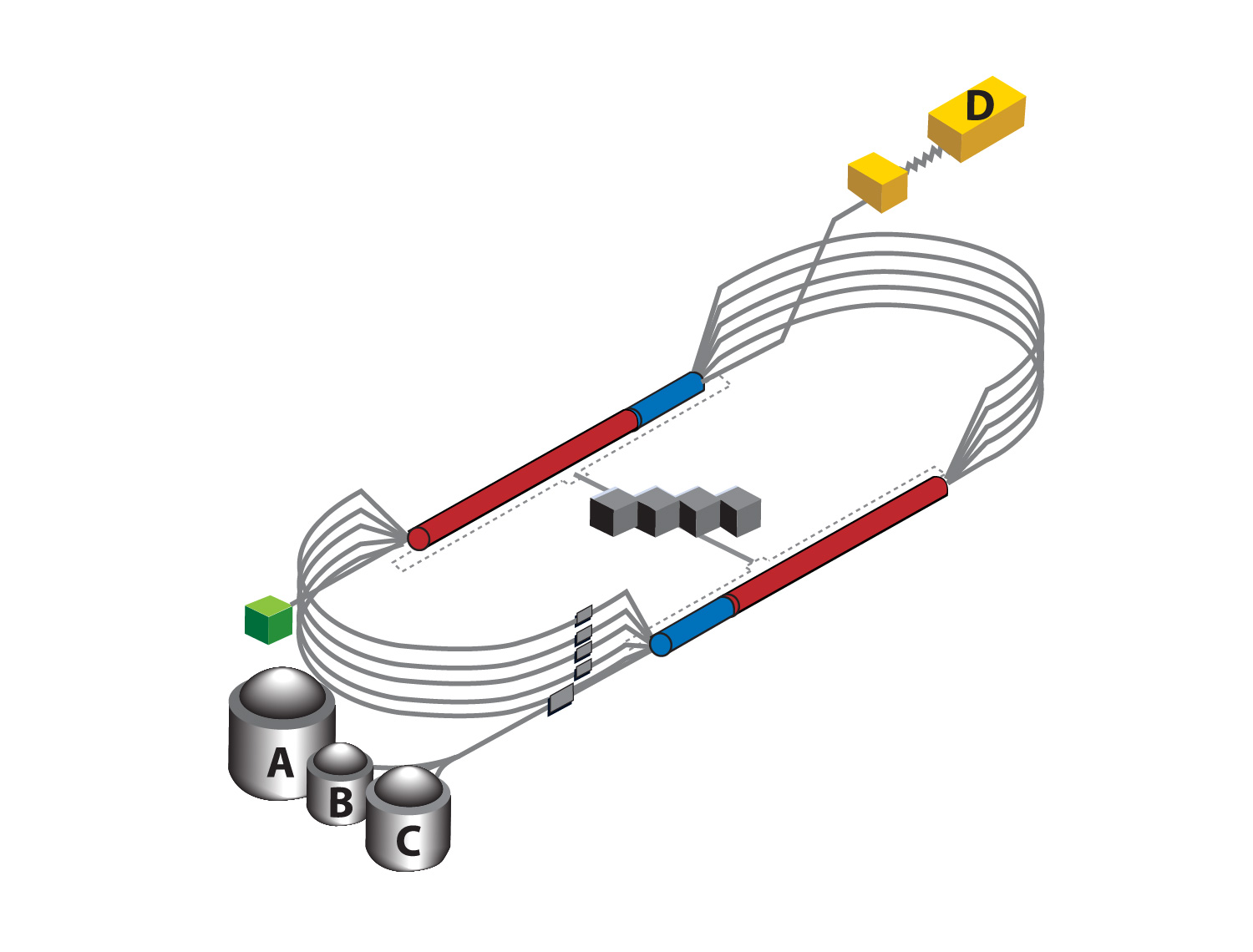}
    \caption{CEBAF layout.  Electron beam is injected at 123 MeV/c from green box at left.  The red/blue cylinders represent the original and newer complement of SRF cavities providing 1100 MeV momentum gain per linac.  At the end of the linac is a series of magnets, including septa, which separate different momenta beams into five arcs of physical radius 80.6 m.  These are termed spreaders.  At the end of each arc is a mirror image set of magnets which recombine the five momentum beams so they may enter the next linac.  In the lower left of the figure is a stack of grey boxes.  These represent RF separators which are followed by the thin septa of Figure 1 to extract beam to halls A, B or C.}
    \end{figure}

Since the spreaders and recombiners are mirror images, albeit with slight current differences due to synchrotron radiation losses in the arcs, septum magnets on either end of each arc are powered with one supply.  About 600 m of cable is required to join them to the supply so the high DC current of the BNL septum \cite{3} is not practical.  CEBAF septa have multiple small conductors with small water holes to achieve the needed AmpTurns.  As a result, in spite of extensive treatment given to the low conductivity water used to cool the magnets, clogging is an issue and a weak acid must be used every few years to clean the cooling passages.  

\section{Septum design work for 22 GeV}

As mentioned in the abstract, there is an exercise underway to explore the possibility of increasing the energy to Halls A/B/C from 11 GeV to 22 GeV.  The highest momentum (lowest physical) arcs in Figure 2 would be replaced with permanent magnets providing an FFA lattice which can recirculate six beams.  The first four passes would reuse existing electromagnets for the most part.  This will require a new septum design which can accept beams of twice the momentum.  Given the limited space available in the existing tunnel the author looked to a conductively cooled superconducting design using MgB2 or a similar conductor with adequate current density at 20 K. \cite{4}  Based on prior experience it was assumed that the cryostat would extend 15 mm on all sides from the superconductor.  The section of one of the first attempts is shown in Figure 3. Another option for the superconductor is Bi-2223 which has adequate engineering Jc at 20 K. \cite{5} A useful HTSC reference is \cite{6}.

\graphicspath{{Images/}}
\begin{figure}[htp]
    \centering
    \includegraphics[width=8cm]{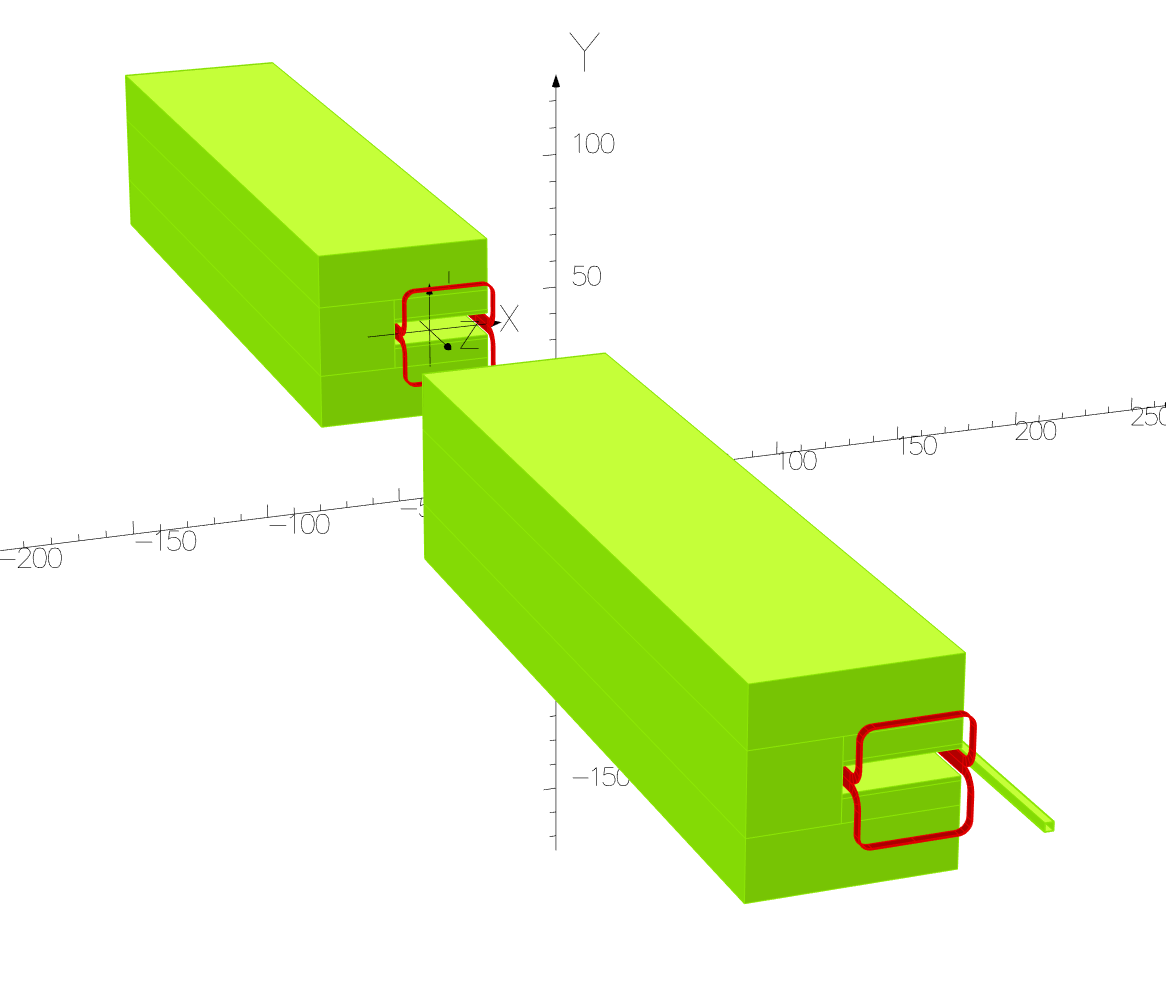}
    \caption{Seven beams with different momenta enter the rear magnet.  The front magnet is offset 7 cm to the left so the lowest momentum beam enters the steel tube on the far right which shields it from the stray field outisde the current sheet in the second magnet.  The 15 mm gap all around the coil for the cryostat increases stray flux.  Symmetric bedsteads are used in the coil model to provide one symmetry which reduces the size of the FEM model.  In a real magnet there would be only one bedstead, full height.  The Z gap between the two magnets is 150 cm to accommodate the needed cryocoolers.  The back leg is the same size as the current sheet as superconductor is assumed.  Current density 206 $A/mm^2$, 1.05 T at the conductor. }
        \end{figure}
        
The stray field resulting from the gap between conductor and steel necessitated by the cryostat led me to look at closing the pole gap from 90 mm to 40 mm.  Vacuum vessel wall thickness on the width of the pole is 6 mm for acceptable deflection due to atmospheric pressure.  This reduced the required current density to 100 $A/mm^2$ so the demands on the superconductor are much reduced.  This variation is shown in Figure 4. 

\graphicspath{{Images/}}
\begin{figure}[htp]
    \centering
    \includegraphics[width=8cm]{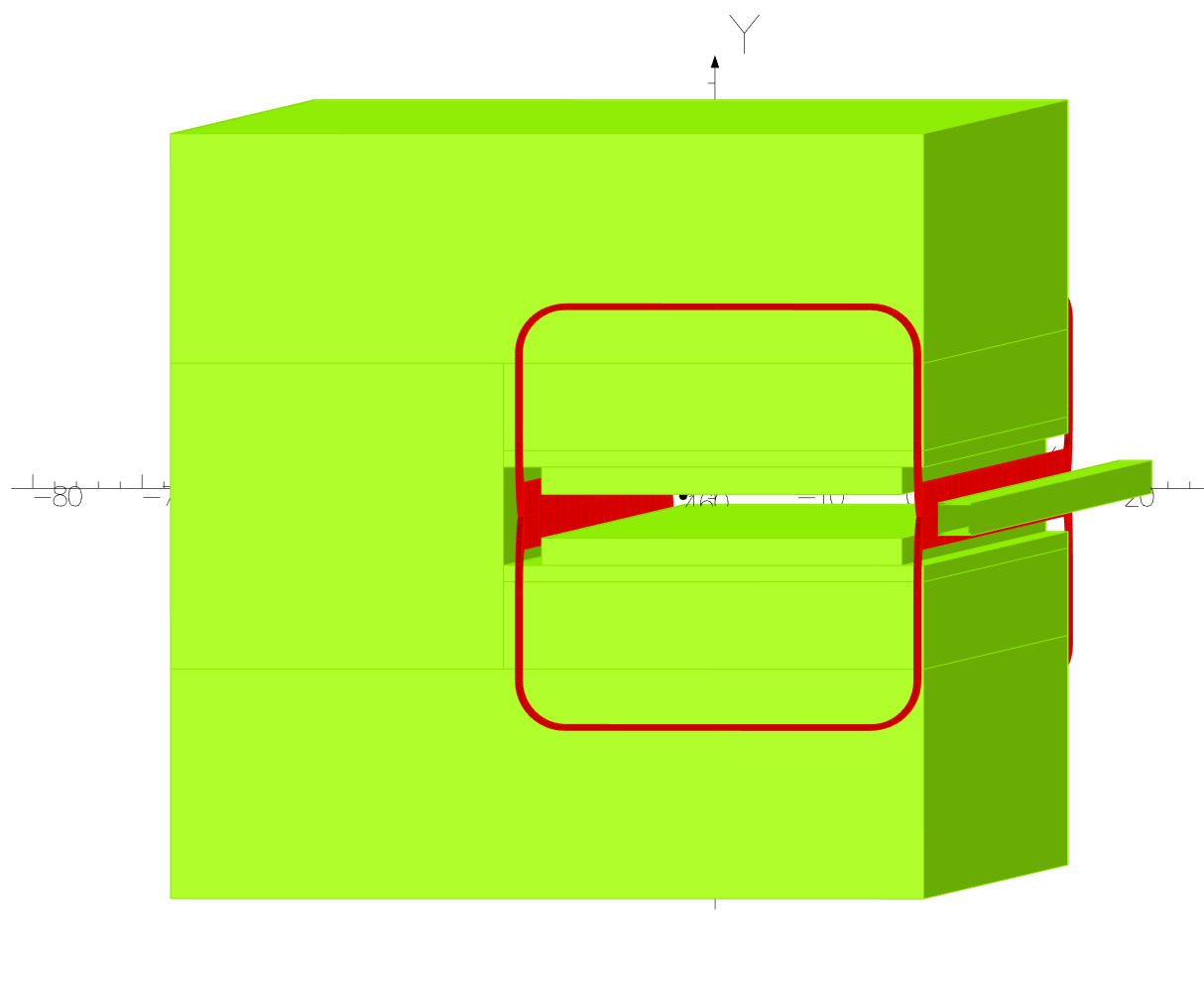}
    \caption{Perspective view of downstream magnet and shield tube with pole separation 40 mm.  15 mm allowance for cryostat surrounds the coil.  }
    \end{figure}

 For amusement because the author is peripherally involved with PERLE \cite{7}, the model was run with $300 A/mm^2$.  This allows for deflecting a 48 GeV electron beam, Figure 5, with the steel ~2.5 T.    

\graphicspath{{Images/}}
\begin{figure}[htp]
    \centering
    \includegraphics[width=14cm]{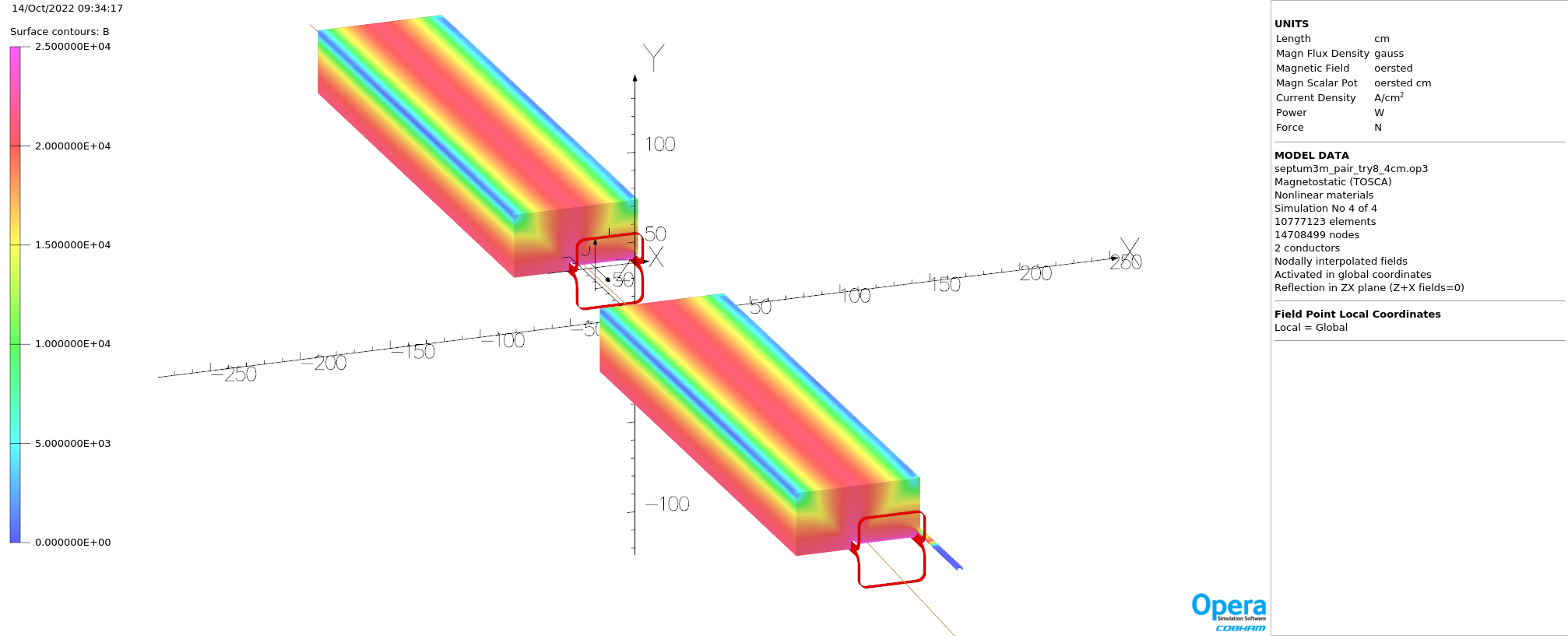}
    \caption{Model with 40 mm pole gap and J = 300 $A/mm^2$.  Track of 48 GeV electron is barely visible. }
    \end{figure}
 
Since the current density in the superconductor had dropped to 100 $A/mm^2$ with the 40 mm pole gap and significant volume was used for the cryostat, copper was attempted.  The major insight here is that the conductor does not have to be placed completely between the poles as long as the steel is not saturated and the passing beam is within a carbon steel beam tube.  The coil and cryostat in the previous model was 35 mm wide by 45 mm half-height.  The gap in the steel for the back leg is 37 mm wide by 45 mm high.  For the current sheet the outer 15 mm of cryostat was outside the steel.  The steel was altered in the model to allow both notches to be 32 mm wide by 45 mm high.  Luvata \cite{8} has dies for 10 mm square conductor with 7.5 mm hole, 1 mm corner radius.  Each of the two bedsteads would be three turns wide by four turns high.  They would be wound as three single pancakes four turns high per coil.  Each pancake need have only one water connection if 25 bar cooling water were available.  The total number of water connections would then be only six versus twenty-four in the existing 3000 m septum, about which more below.  One could also have four pancakes with three turns each to lower required water pressure, ~14 bar.  Water flow rate 0.3 l/s assumed in both cases.  Required 14000 AmpTurns yields 1166 A, about half-again higher than the largest DC current now used in CEBAF \cite{2} but still low compared to the BNL magnet \cite{3}. 

\graphicspath{{Images/}}
\begin{figure}[htp]
    \centering
    \includegraphics[width=8cm]{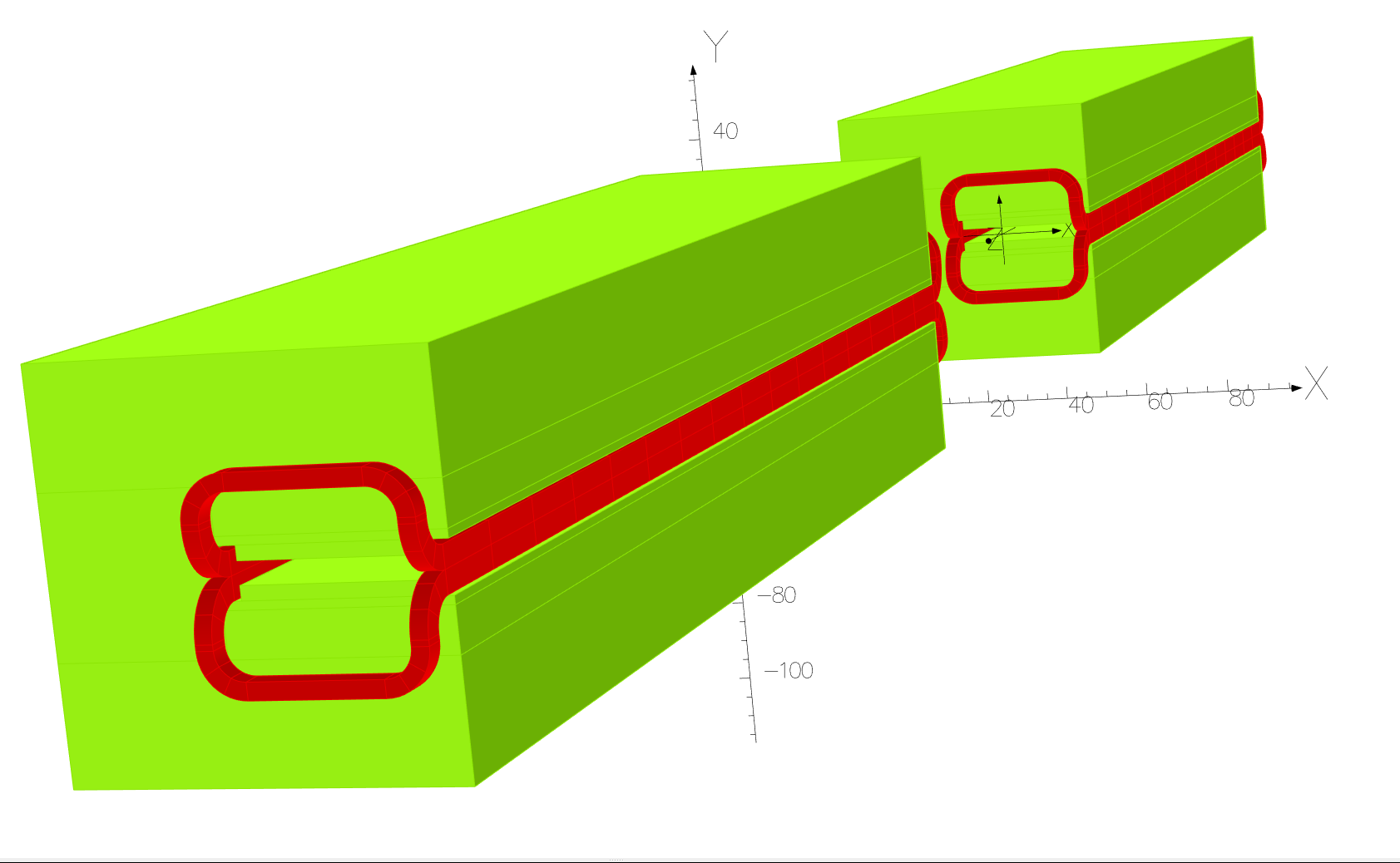}
    \caption{System with same steel as Figures 4 and 5 but copper coils.  Shield tube for passing beam removed for clarity. }
    \end{figure}

Fourier harmonics for the three variations discussed above were calculated along the orbits of the different momentum beams which will see these magnets in the 22 GeV concept.  The 40 mm gap provides better field uniformity than the 90 mm gap with either superconducting or conventional coils.  See appendix A for the values in “units”, 1E-4 of dipole field.  These are calculated along trajectories propagated through the model by Opera \cite{11} at 2.5 mm intervals.  Circles of 1 cm radius are placed perpendicular to the Z axis at each point and By calculated on 60 points on each circle.  The Fourier Fit function is then applied to calculate the harmonics.  Values for all the circles on an orbit are summed and divided by four to get centimeters. Figure 7 shows the orbits through one of the spreaders as presently defined; exact orbits will change. 

\graphicspath{{Images/}}
\begin{figure}[htp]
    \centering
    \includegraphics[width=15cm]{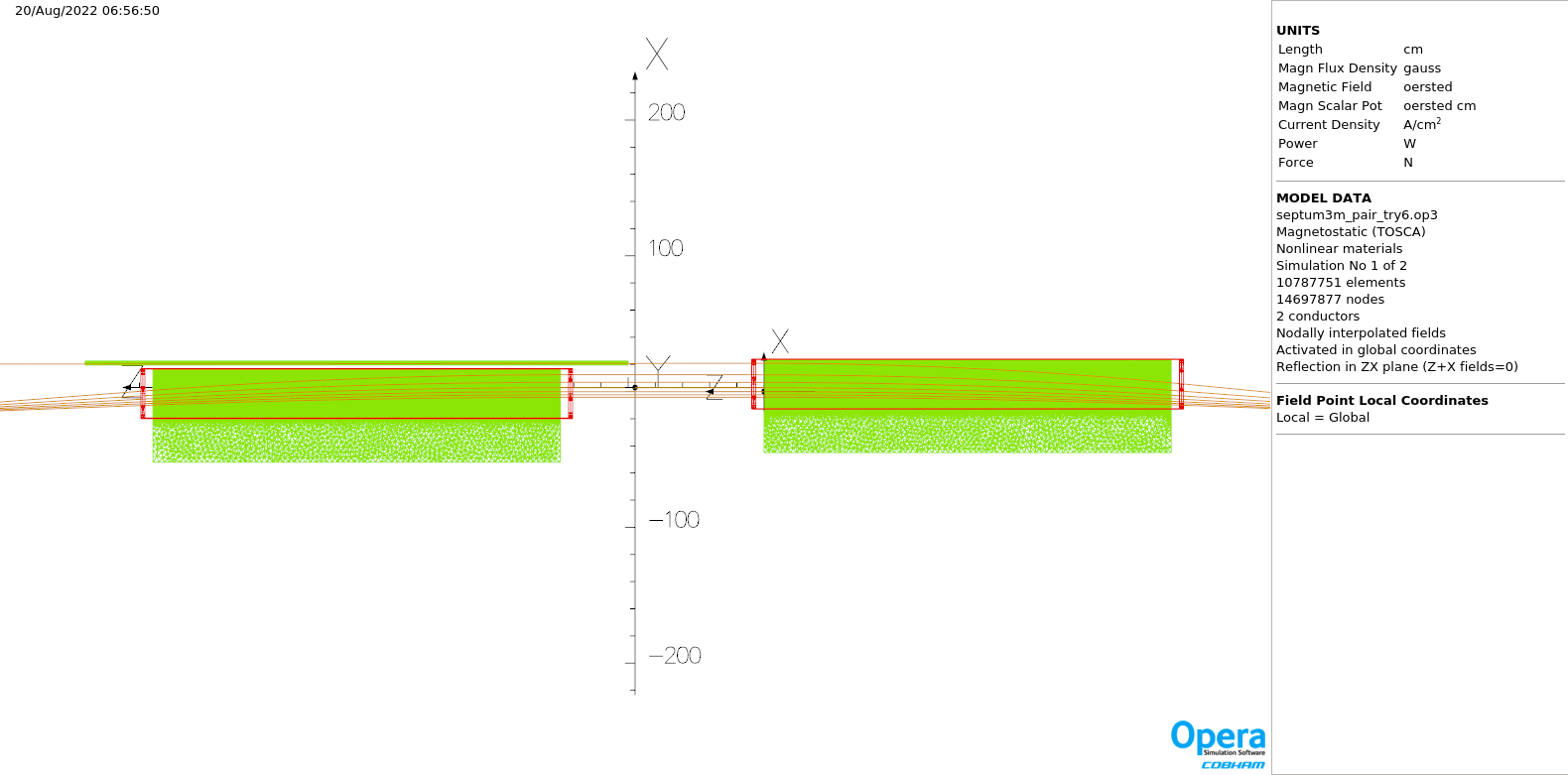}
    \caption{Seven orbits through the NE spreader concept at 22 GeV. }
    \end{figure}

\section{Modifying an existing septum magnet}

CEBAF includes eight septa with nominal length 3000 mm.  Steel length is 2982 mm.  The vacuum chamber is 3349 mm and there is 33.6 mm of stainless steel between the beam paths in the entry flange.  The coil is built in two halves.  Each has twelve turns arranged six high by two wide of Luvata 8327 \cite{8}.  This conductor is 4 mm by 6 mm with 0.8 mm corner radius and 2.5 mm hole.  Copper area of one conductor 17.08 $mm^2$.  Current density 48 $A/mm^2$.  In the front (current sheet) leg the coil pack including insulation is ~ 26 $mm^2$.  In the back leg the same conductor is brazed into 13.5 mm by 12.4 mm copper bus so the net copper area is 160.63 $mm^2$ and current density 5.07 $A/mm^2$.  There is no supplemental cooling in the back leg, only the 2.5 mm hole.  There are individual water paths for each turn.

These coils are prone to clogging.  Two episodes in the last year totalled 152 hours of down time.  With beam recovery, about one week of physics was lost.  Additional labor is required to check flow during scheduled accelerator downs and then flushing to open clogged passages.  The current sheet and steel are curved with a radius of 40.98 m on the side where the beams propagate while the return steel is straight. 

\graphicspath{{Images/}}
\begin{figure}[htp]
    \centering
    \includegraphics[width=15cm]{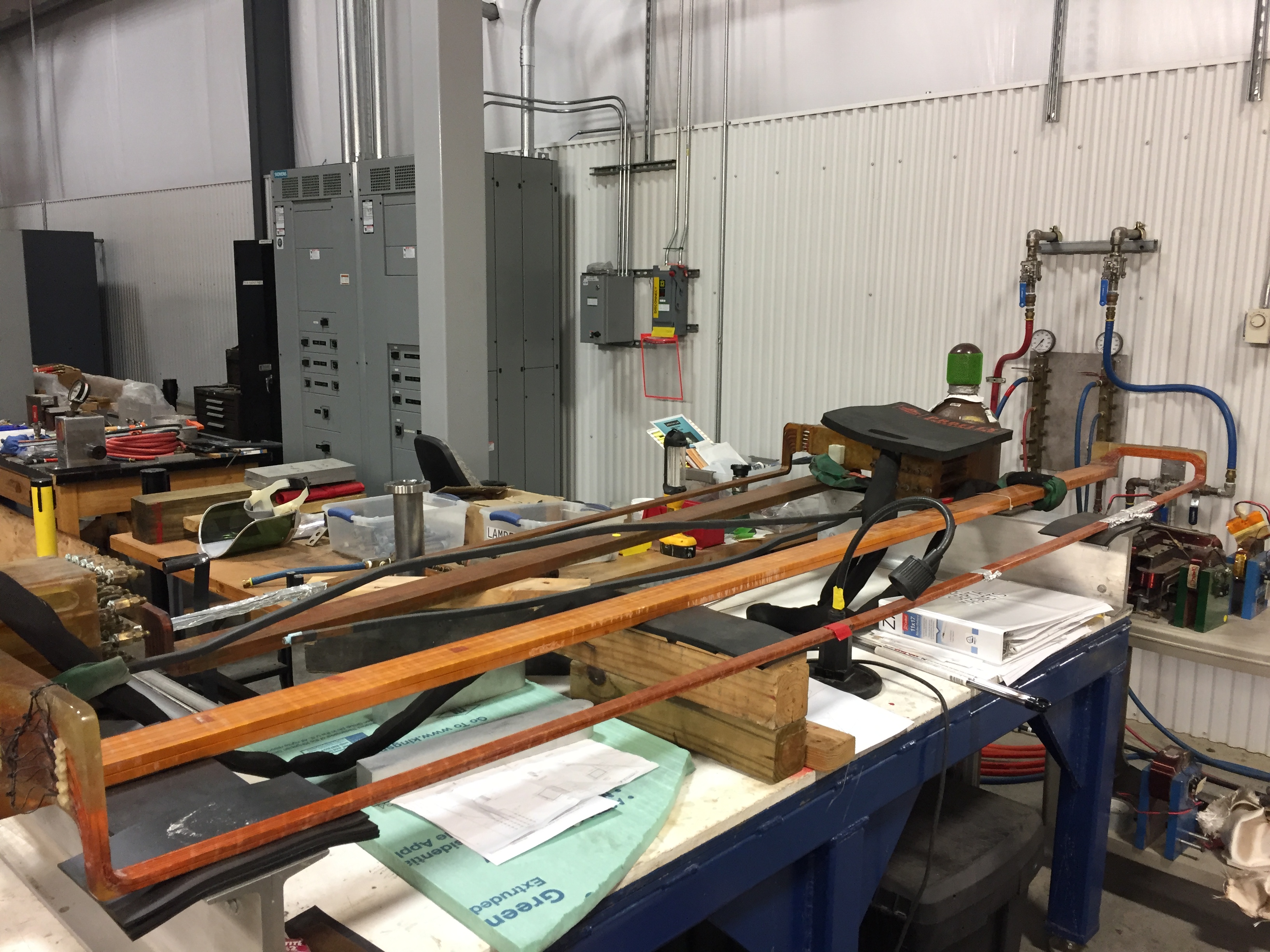}
    \caption{Spare coil assembly.  The smaller set of turns in the foreground will be half the current sheet.  The 41 m radius of this set is not apparent.}
    \end{figure}

\graphicspath{{Images/}}
\begin{figure}[htp]
    \centering
    \includegraphics[width=15cm]{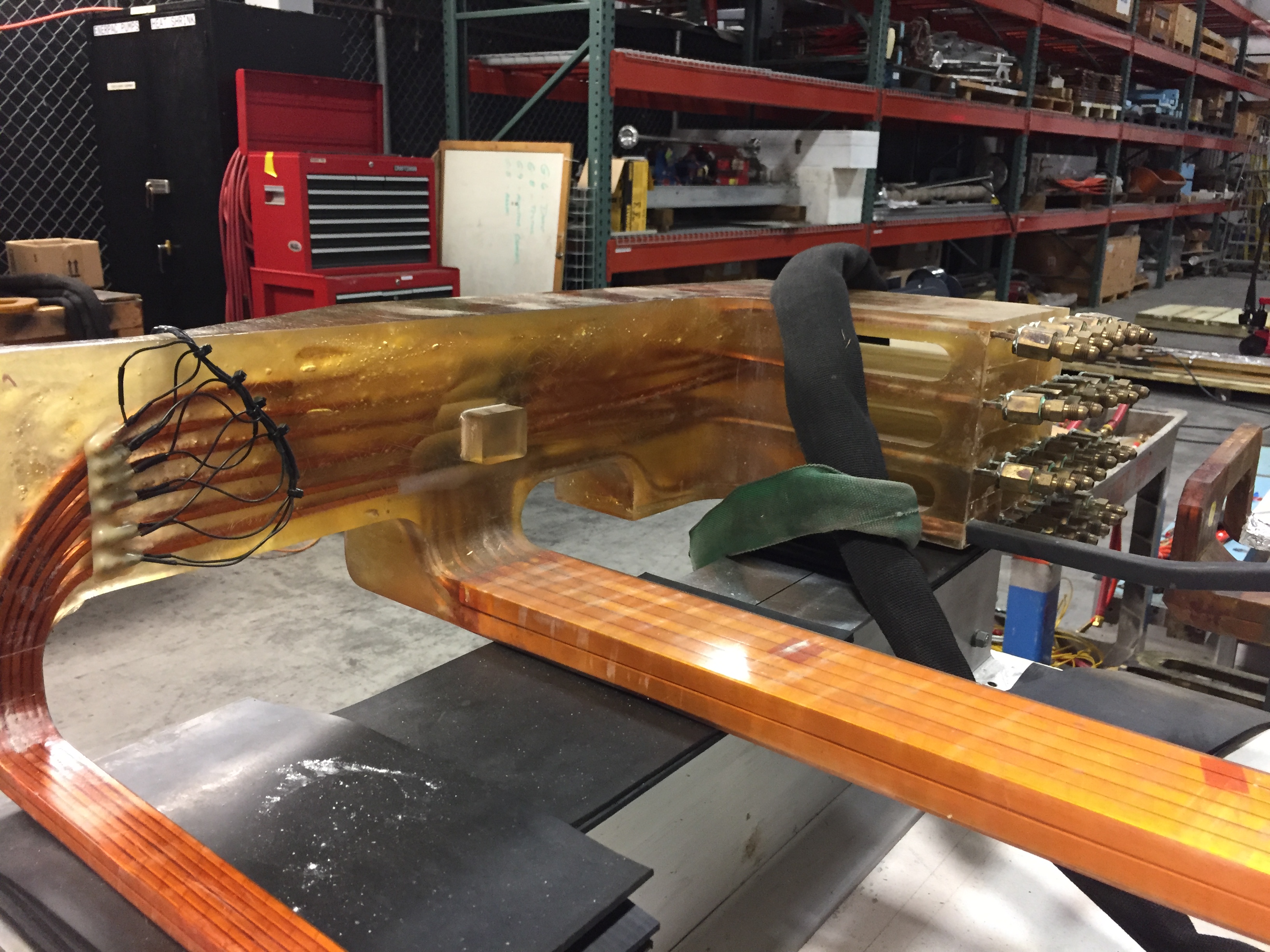}
    \caption{Closeup of end of the coil.  The twelve small turns are at the left.  The black wires lead to temperature sensors epoxied to each turn.  The back leg is in the center.  The water connections are far right.  }
    \end{figure}

Several simplifying assumptions were used in the models shown below.  The steel and current sheet are not curved.  The additional copper on the back leg is not included, only simple bedsteads are used.  Since the region of interest is close to the current sheet the exact current distribution in the back leg is not relevant.  A model without these assumptions for the 2000 mm counterpart to this magnet is discussed in \cite{9}. 

Fields were calculated in Opera \cite{11} along straight lines offset 9 mm from either side of the coils in the models, 42 mm total separation at the entrance to the magnet.  The beam within the bore of the magnet will obviously not have a straight path but this is a convenient way to compare the two designs. 

\begin{table}[htbp]
\centering
\caption{\label{tab:i} Fields of the models in Figures 10 and 11, straight lines referenced to coil}
\smallskip
\begin{tabular}{c c c c}
\hline
model&$\int$ Bdl 9 mm inside&$\int$ Bdl 9 mm outsidel&outside/inside \%\\
\hline
standard current sheet&2757420 G-cm&23874 G-cm&0.87\%\\
recessed current sheet&2760830 G-cm&1954 G-cm&0.07\%\\
\hline
\end{tabular}
\end{table}

Assume conductor Luvata 8328 \cite{8}, 8 mm by 6 mm, 3.2 mm hole, 0.8 mm corner radius.  
Standard current sheet 19480.3 AT/24T = 812 A, current density in copper 47.5 $A/mm^2$
Recessed current sheet 20252.1 AT/24T = 844 A, current density in copper 24.4 $A/mm^2$
Current density 51.4 \% of that in the original design yielding 0.12 \% higher $\int B\,dl\ $

A marked improvement and much easier to cool.  Coils will cost much less to fabricate.  As shown in the previous section on the 22 GeV design, this idea can be applied to a variety of DC septum magnets.  Not only is the current density about half of that of a classic DC current septum but the field outside the current sheet is less than one tenth of the classical septum.  Two very significant benefits. 

\graphicspath{{Images/}}
\begin{figure}[htp]
    \centering
    \includegraphics[width=15cm]{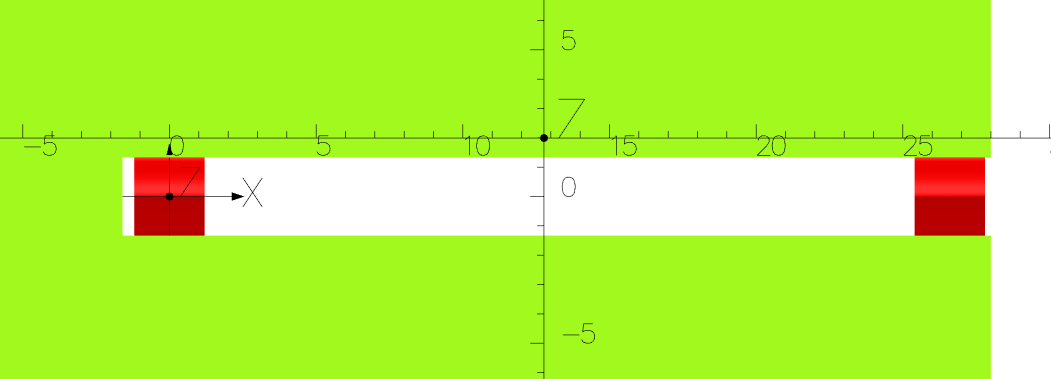}
    \caption{ Section of approximation to existing magnet with the current sheet entirely between the steel.  }
    \end{figure}

\graphicspath{{Images/}}
\begin{figure}[htp]
    \centering
    \includegraphics[width=15cm]{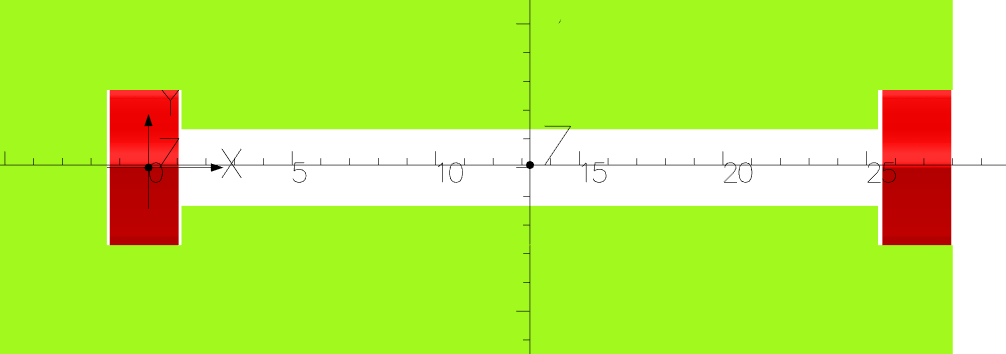}
    \caption{ Section of the proposed modification with half the coil cross-section recessed within the steel.  Overall steel section is the same, only the differences between the two models are shown.  }
    \end{figure}

\section {Conclusion}

A simple insight allows for the reduction in current density in DC current sheet septum by a factor of two while reducing the stray field by a factor of ten.  Harmonic content is not quite as good as with the narrower superconducting coil first assumed for 22 GeV but still acceptable [Appendix A].

\pagebreak

\acknowledgments
This material is based upon work supported by the U.S. Department of Energy, Office of Science, Office of Nuclear Physics under contract DE-AC05-06OR23177.


\appendix
\section{Fourier harmonics expressed in Units (0.01\% of dipole) along multiple orbits for three models of Section 2}
%\begin{table}[htbp]
\centering
\begin{tabular}{c c c c c c c}
\hline
model&energy (MeV)&dipole G-cm&quadrupole&sextupole&octapole&decapole\\
\hline
SC 9 cm gap&8350&-2715798&89.4&27.4&4.8&0.3\\
SC 4 cm gap&8350&-2664172&-77.6&-51.1&-18.7&-2.5\\
Cu 4 cm gap&8350&-2662116&-85.8&-56.9&-21.6&-3.6\\
SC 9 cm gap&9450&-2799159&72.0&22.9&4.3&0.5\\
SC 4 cm gap&9450&-2760532&-45.9&-32.2&-13.9&-3.5\\
Cu 4 cm gap&9450&-2758108&-50.7&-35.5&-15.5&-4.1\\
SC 9 cm gap&10550&-5370076&19.9&6.7&1.4&0.2\\
SC 4 cm gap&10550&-5353327&-5.7&-3.8&-1.9&-0.7\\
Cu 4 cm gap&10550&-5352134&-6.2&-4.2&-2.1&-0.7\\
SC 9 cm gap&11650&-5549161&17.0&5.8&1.2&0.2\\
SC 4 cm gap&11650&-5535545&-4.1&-2.7&-1.3&-0.5\\
Cu 4 cm gap&11650&-5532448&-4.5&-2.9&-1.5&-0.5\\
SC 9 cm gap&12750&-5354037&0.1&0.2&0.0&0.0\\
SC 4 cm gap&12750&-5355188&-0.5&0.0&0.0&0.0\\
Cu 4 cm gap&12750&-5354043&-0.5&0.0&0.0&0.0\\
SC 9 cm gap&13850&-5535208&0.0&0.2&0.0&0.0\\
SC 4 cm gap&13850&-5537246&-0.5&0.0&0.0&0.0\\
Cu 4 cm gap&13850&-5534151&-0.5&0.0&0.0&0.0\\
SC 9 cm gap&14950&-5354066&-1.2&0.4&-0.1&0.0\\
SC 4 cm gap&14950&-5355209&-0.3&0.0&0.0&0.0\\
Cu 4 cm gap&14950&-5354040&-0.3&0.0&0.0&0.0\\
SC 9 cm gap&16050&-5535609&-1.4&0.4&-0.1&0.0\\
SC 4 cm gap&16050&-5537395&-0.3&0.0&0.0&0.0\\
Cu 4 cm gap&16050&-5534283&-0.3&0.0&0.0&0.0\\
SC 9 cm gap&17150&-5356328&-4.4&1.5&-0.3&0.1\\
SC 4 cm gap&17150&-5355097&0.0&-0.2&0.1&0.0\\
Cu 4 cm gap&17150&-5353931&0.0&-0.3&0.1&0.0\\
SC 9 cm gap&18250&-5539595&-6.6&2.2&-0.5&0.1\\
SC 4 cm gap&18250&-5537193&0.4&-0.4&0.2&-0.1\\
Cu 4 cm gap&18250&-5534050&0.5&-0.6&0.2&-0.1\\
SC 9 cm gap&19350&-5363035&-13.1&4.3&-0.9&0.1\\
SC 4 cm gap&19350&-5354307&2.1&-1.7&0.8&-0.3\\
Cu 4 cm gap&19350&-5352955&2.6&-2.0&0.9&-0.3\\
SC 9 cm gap&20450&-5553188&-22.9&7.3&-1.4&0.2\\
SC 4 cm gap&20450&-5534395&7.6&-5.5&2.4&-0.7\\
Cu 4 cm gap&20450&-5530659&9.1&-6.4&2.8&-0.8\\
SC 9 cm gap&21550&-5377752&-31.1&9.8&-1.9&0.2\\
SC 4 cm gap&21550&-5350096&13.2&-9.2&3.8&-0.9\\
Cu 4 cm gap&21550&-5348012&15.5&-10.6&4.4&-1.1\\
\hline
\end{tabular}
%\end{table}

\end{document}